 \documentclass[preprint, prb, showpacs, superscriptaddress]{revtex4}

\usepackage{graphicx}

\begin{document}

\title{Dirac Semimetal and Topological Phase Transitions in $A_3$Bi
  ($A$=Na, K, Rb)}





\author{Zhijun Wang}
\affiliation{Beijing National Laboratory for Condensed Matter
  Physics, and Institute of Physics, Chinese Academy of Sciences,
  Beijing 100190, China;\\}

\author{Yan Sun}
\author{Xing-Qiu Chen}
\author{Cesare Franchini} 

\affiliation{Shenyang National Laboratory for Materials Science,
  Institute of Metal Research, Chinese Academy of Sciences, Shenyang
  110016, China}
  
 \author{Gang Xu}
 \author{Hongming Weng}\email{hmweng@aphy.iphy.ac.cn}
 \author{Xi Dai}
 \author{Zhong Fang}\email{zfang@aphy.iphy.ac.cn}

\affiliation{Beijing National Laboratory for Condensed Matter
  Physics, and Institute of Physics, Chinese Academy of Sciences,
  Beijing 100190, China;\\}


\date{\today}

\begin{abstract}
  The three-dimensional (3D) Dirac point, where two Weyl points overlap in momentum space, is
  usually unstable and hard to realize.  Here we show, based on the
  first-principles calculations and effective model analysis, that
  crystalline $A_3$Bi ($A$=Na, K, Rb) are Dirac semimetals with bulk
  3D Dirac points protected by crystal symmetry. They possess
  non-trivial Fermi arcs on the surfaces, and can be driven into
  various topologically distinct phases by explicit breaking of
  symmetries. Giant diamagnetism, linear quantum magnetoresistance,
  and quantum spin-Hall effect will be expected for such compounds.
\end{abstract}

\maketitle

\section{introduction} \label{introduction}

The topological consideration of effective relativistic quantum field
theory in three dimensional (3D) momentum space allow us to classify
quantum vacuum into three distinct classes, namely, that with gap,
fermi surfaces (FSs) and Weyl points.\cite{standardmodel,volovik-book} The
topological classes with fermi surface or gap are well known in
condensed matters as metals or insulators. The insulators have been
further classified into topologically trivial and non-trivial
insulators.~\cite{TIreview,TIreview-2,TIclass-1} For the later case,
there have been many candidate materials being proposed and some of
them confirmed by experimental
observations.~\cite{TIreview,TIreview-2,TIclass-1,zhangnphy,Xia,Chen}
The class of compounds with Weyl fermi points are relatively rare,
however the $A$ phase of $^3$He~\cite{He3} and some recent
proposals~\cite{wan, HgCrSe, multilayerTRB, nodalsemimetal} have
suggested various possibilities. In addition to Weyl fermi points, as
we will demonstrate in this paper, the fermi surfaces can be further
classified~\cite{standardmodel,volovik-book} and each of them can be
realized in condensed matters.  On the other hand, given those known
realizations of distinct topological states, it is yet challenging to
have a well controlled example near the phase boundary so that various
topological phase transitions can be studied systematically within one
system. This challenge becomes further relevant because the vacuum of
the Standard Model is regarded to be at the phase boundary with marginal
Fermi points (MFP), which is composed of two overlapping Weyl points
with opposite chirality (or topological charge), i. e., forming
3D massless Dirac points.~\cite{standardmodel,volovik-book} The
Weyl points with opposite chirality are stable topological objects
only when they are separated. If they meet in momentum space, their
topological charges may cancel each other and open a gap. In
principle, we may accidentally obtain 3D Dirac points by finely tuning
chemical composition or spin-orbit coupling (SOC)
strength.~\cite{Murakami,ZhangW,Hasan-phaseT,Ando-phaseT}
Unfortunately, such realizations are too fragile and hard to
control. In fact, in the presence of crystal symmetry, the 3D Dirac
points can be protected and stabilized, as has been discussed in
Ref. 18
and will be addressed in this paper.

Here, we report that the 3D Dirac semimetal (or MFP) state can
be achieved in a simple stoichiometry compound $A_3$Bi ($A$=Na, K,
Rb), where the low energy states form an effective 3D massless gas of
Dirac fermions, being different from those in graphene (2D Dirac
fermion) and Weyl semimetals (non-overlapping Weyl points). This state
is located at the phase boundary, and is stabilized by the crystal
symmetry. It can be driven into various topologically distinct phases
by explicit breaking of symmetries, and thus provides us a nice
example for the systematical studies of topological phase
transitions. In addition, the state itself is topologically
non-trivial in the sense that it has Fermi arcs on the surfaces, and
it show giant diamagnetism and quantum magnetoresistance (MR) in the
bulk. It can also show the quantum spin Hall effect in its quantum-well
 (or thin-film) structure. We will start from the structure and
methods in Sec. II, present the main results in Sec. III, and
finally conclude in Sec. IV.

\begin{figure}[tbp]
\includegraphics[height=65mm]{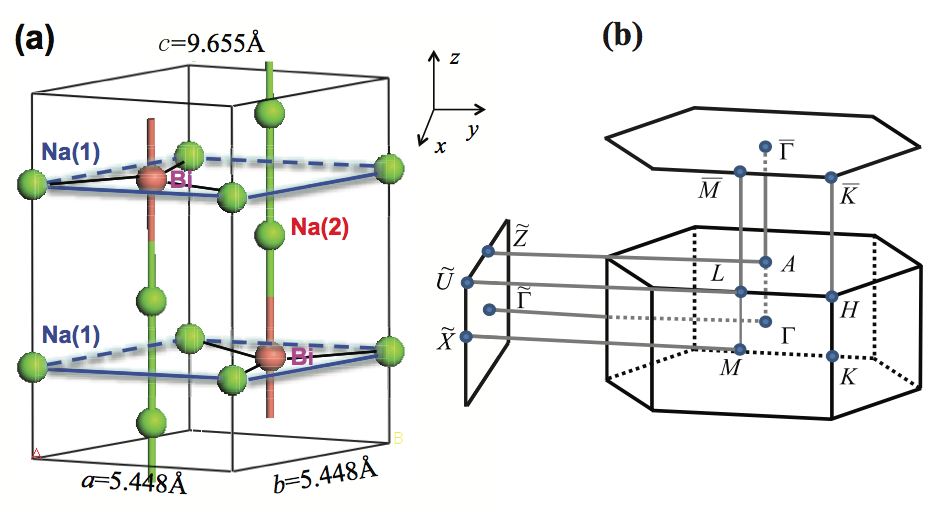}
\caption{(Color online) (a) Crystal structure of Na$_3$Bi with
  $P6_3/mmc$ symmetry. Na(1) is at $2b$ position
  $\pm$(0,0,$\frac{1}{4}$), and Bi is at $2c$ position
  $\pm$($\frac{1}{3}$,$\frac{2}{3}$,$\frac{1}{4}$). They form
  honeycomb lattice layers. Na(2) is at $4f$ position
  $\pm$($\frac{1}{3}$, $\frac{2}{3}$,$u$) and
  $\pm$($\frac{2}{3}$,$\frac{1}{3}$,$\frac{1}{2}+u$) with $u$=0.583,
  threading Bi along the $c$ axis. (b) Brillouin Zone of bulk and the
  projected surface Brillouin Zones of (001) and (010) plane.}
\end{figure}

\section{Crystal Structure and Methodology} \label{Methodology}

Among the alkali pnictides $A_3B$ ($A=alkali$ metal, $B$=As, Sb or
Bi), $A_3$Sb is well known for its application as a photocathode
materials,~\cite{A3Sb-zunger} but the physical properties of $A_3$Bi
are not widely studied.~\cite{A3Bi-elec} Both Li$_3$Bi and Cs$_3$Bi
crystallize in cubic $Fm\bar{3}m$ structure, while Na$_3$Bi, K$_3$Bi and
Rb$_3$Bi are in hexagonal $P6_3/mmc$ phase (or $D_{6h}^4$, shown in
Fig.1),~\cite{A3B-struc} which are our main focus here.  Taking
Na$_3$Bi as an example, in this structure,~\cite{Na3Bi-struc} there
are two nonequivalent Na sites (Na(1) and Na(2)). Na(1) and Bi form
simple honeycomb lattice layers which stack along the $c$ axis, while
Na(2) atoms are inserted between the layers, making connection with Bi
atoms. From the ionic picture, due to the closed-shell configuration
where the number of valence electrons (3$\times$Na-$s^1$+Bi-$p^3$) is
equal to six, we may expect a semiconducting nature of these compounds,
similar to Na$_3$Sb.~\cite{Na3Sb} However, they are in fact different.

To explore the electronic properties of $A_3$Bi, we performed band-structure
 calculations based on the plane-wave ultra-soft
pseudopotential method, using the generalized gradient approximation
(GGA) for the exchange-correlation functional.  The calculations based
on hybrid function (HSE)~\cite{HSE06} are further supplemented to
check the band gap.  The cutoff energy for wave-function expansion is
25 Ry, and the $k$-point sampling grid is 12$\times$12$\times$6. The
experimental lattice parameters~\cite{Na3Bi-struc} are used in
calculation and convergency is checked with the above settings. The
projected surface states are obtained from the surface Green's
function of the semi-infinite system, similar to the method used for
Bi$_2$Te$_3$ and Bi$_2$Se$_3$.~\cite{zhangnphy,ZhangW} For this
purpose, the maximally localized Wannier functions from the
first-principles calculations have been constructed.

\section{Results and discussions} \label{result}

\subsection{Electronic Structures: Fermi points and Fermi arcs}

The calculated electronic structures shown in Fig. 2 suggest that the
valence and conduction bands are dominated by Bi-$6p$ and Na-$3s$
states. Very close to the Fermi level, the top valence band is mostly
from Bi-$6p_{x,y}$ states, while the conduction band with very strong
dispersion is mostly from Na(1)-$3s$ states. All these pictures are
similar to that of Na$_3$Sb,~\cite{Na3Sb} but with the key difference that
at the $\Gamma$ point, the Na-$3s$ band is lower than Bi-$6p_{x,y}$ by
about 0.3 eV, and it is further enhanced to be 0.7 eV in the presence
of SOC, resulting in a metal with an inverted band structure, rather than
the normal narrow gap semiconductor like Na$_3$Sb.~\cite{Na3Sb} The
band inversion is mostly due to the heavier Bi, which has higher
$6p$ states and larger SOC compared to Sb. Considering the possible
underestimation of the band gap by GGA, the band inversion can be further
confirmed by the following evidences: (1) calculation using hybrid
functional HSE gives a band inversion around 0.5 eV, still reasonably
strong; and (2) the earlier calculations for normal semiconductor
K$_3$Sb~\cite{K3Sb} suggest that its experimental gap can be
reasonably reproduced by GGA.  With the same method, we calculate the
band structures for K$_3$Bi and Rb$_3$Bi and find that the band
inversions are 0.33 and 0.42 eV, respectively.  Because of the
similar band structures and the same outcomes of the analysis, we
mainly investigate Na$_3$Bi for details in the following.

\begin{figure}[tbp]
\includegraphics[height=125mm]{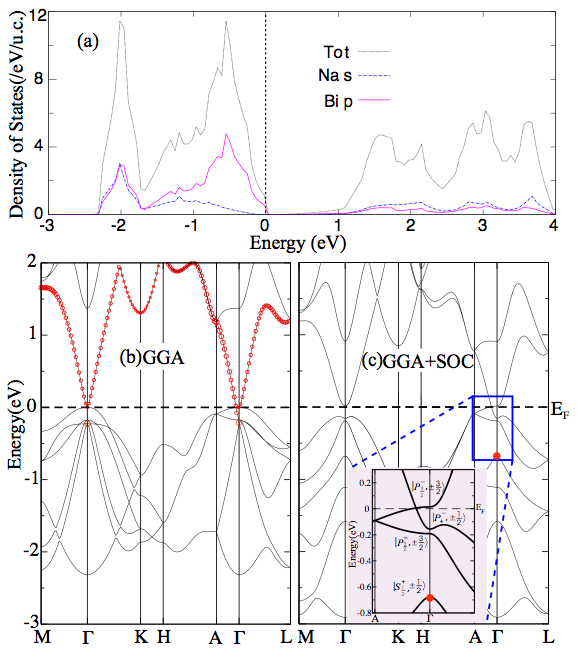}
\caption{(Color online) The calculated electronic structures of
  Na$_3$Bi. (a) The total and partial density of states. (b) and (c)
  are the band structures without and with spin-orbit
  coupling, respectively. The red circles indicate the projection to the Na-3$s$
  states. The orbital characters of wave-functions at $\Gamma$ point
  are labeled in the inset (see Sec. III.B for details).}
\end{figure}

Having the inverted band structure, however, Na$_3$Bi is not gapped,
different from topological insulators like Bi$_2$Te$_3$ and
Bi$_2$Se$_3$.~\cite{zhangnphy} It is a semi-metal with two nodes
(band-crossings) exactly at the fermi level (Fig. 2). In other words, its
fermi surface consists of two isolated fermi points, which are located
at (0, 0, $k_z^c$$\approx$$\pm$0.26$\times \frac{\pi}{c}$) along the
$\Gamma$-$A$ line. Since both time reversal and inversion symmetries are
present, there is four-fold degeneracy at each fermi point, around
which the band dispersions can be linearized, resulting in a 3D
massless Dirac semimetal. It is different from that in graphene not only in
dimensionality, but also in its robustness, because the fermi points
here survive in the presence of SOC. This fact also makes a difference
from other proposals.~\cite{Kino,Pickett}

\begin{figure}[tbp]
\includegraphics[height=95mm]{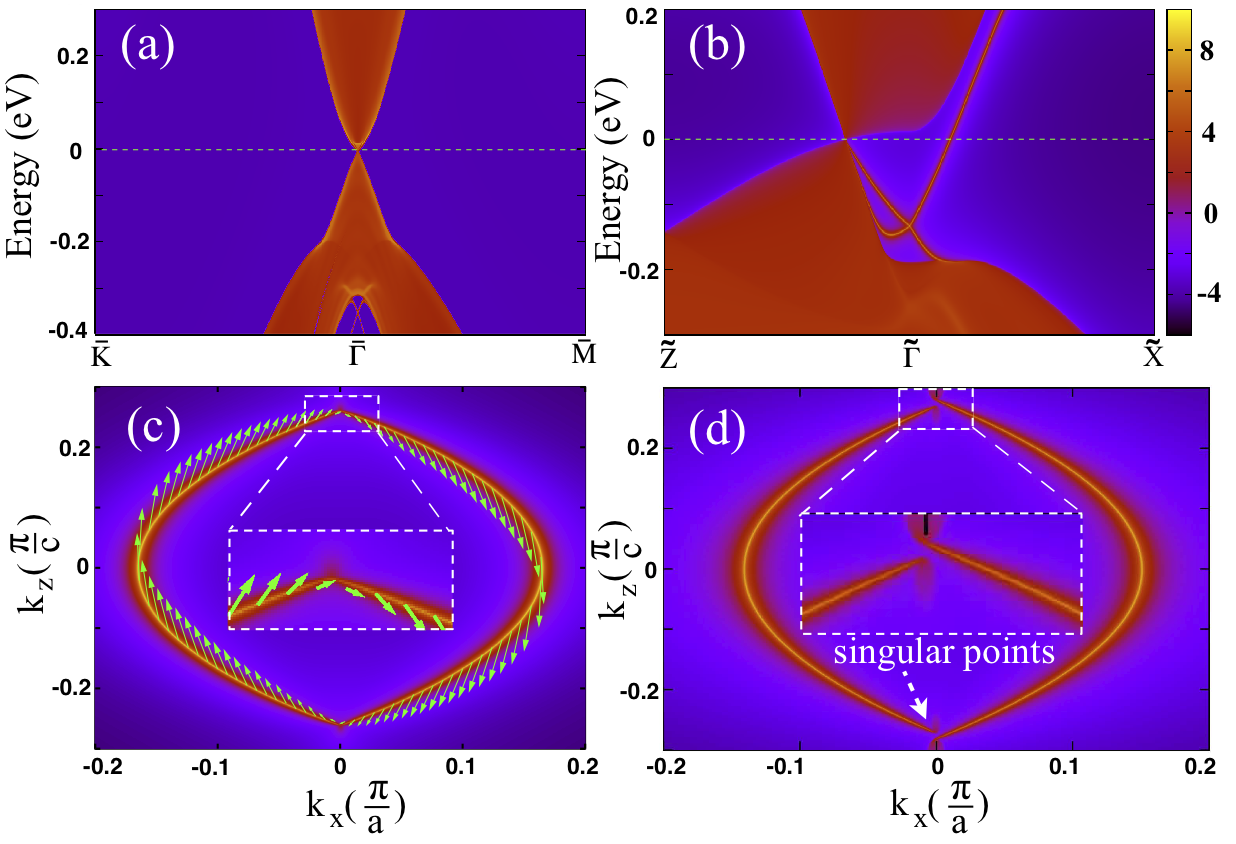}
\caption{(Color online) The projected surface states and their Fermi
  surfaces of Na$_3$Bi. (a) and (b) The projected surface density
  of states for [001] and [010] surfaces, respectively. (c) The Fermi
  surfaces (Fermi arcs) and their spin texture (in-plane component) for
  the [010] surface states. (d) The Fermi arcs of the [010] surface
  obtained from the fitted effective Hamiltonian with additional exchange
  field $h_1$=6 meV (see Sec. III.C for details). The discontinuity
  around the singular Fermi points becomes now obvious (enlarged in
  the insets).}
\end{figure}

The 4$\times$4 Dirac fermion here is massless because the two bands
which cross each other along the $\Gamma$-$A$ line belong to different
irreducible representations under three-fold rotational
symmetry. Breaking of this symmetry will introduce interaction between
them and make the system insulating. For example, 1\% compression
along the $y$ axis will open up a gap of $\approx$5.6 meV. This insulating
state, however, is topologically non-trivial with
$Z_2$=1~\cite{TIreview,TIreview-2} due to the inverted band structure
around the $\Gamma$ point. This fact makes Na$_3$Bi unique, because
both bulk 3D Dirac points and non-trivial surface states (a single
pair) should coexist (see Fig. 3) as long as the crystal symmetry
stands. Furthermore, the surface states are different from that of
topological insulators,~\cite{zhangnphy} in the sense that their Fermi
surfaces has fermi arc structures. As shown in Fig. 3(b) for the [010]
surface of stoichiometry Na$_3$Bi, although the entire fermi surface
is closed, its derivative and fermi velocity are ill defined at the
two singular points (corresponding to the projection of bulk Dirac
points to the surface). The spin texture of surface states has a helical
structure (also similar to topological insulators), but the magnitude
of spin vanishes at the singular points. This kind of fermi surfaces
has never been found before, and it can be understood following the
discussions for Weyl semimetal.~\cite{wan,HgCrSe} If we split the
4$\times$4 Dirac point into two separated 2$\times$2 Weyl points in
momentum space by breaking time reversal or inversion
symmetry,~\cite{multilayerTRB,nodalsemimetal} the fermi surface of
surface states will also split into open segments which are fermi arcs
discussed in Weyl semimetal (as shown in Fig. 3(d)).~\cite{wan,HgCrSe}
All these characters in contrast to conventional metals and
topological insulators should be experimentally measurable by modern
angle-resolved photoemission spectroscopy technique. Our further calculations for K$_3$Bi and Rb$_3$Bi
suggest that they have qualitatively the same physics as Na$_3$Bi
does.

\subsection{Effective Hamiltonian}


The low energy effective hamiltonian is derived from the theory of
invariants in a similar way as for Bi$_2$Se$_3$, Bi$_2$Te$_3$ and
Sb$_2$Te$_3$.~\cite{cxliu} The first-principles calculations indicate
that the wave-functions of low energy states at the $\Gamma$ point are
mostly from the Na-3$s$ and Bi-6$p_{x,y,z}$ orbitals. For the low
energy Na-3$s$ stats at $\Gamma$, about 65\% of them are from Na(1)-3$s$
and 35\% are from Na(2)-3$s$. Since the system has inversion symmetry, we
can start from the bonding and antibonding states of the above relevant
orbitals with definite parity:

\begin{eqnarray*}
  |S^{\pm}\rangle=\frac{1}{\sqrt{2}}(|Na,s\rangle \pm |Na^{'},s\rangle), \\
  |P^{\pm}_{\alpha}\rangle=\frac{1}{\sqrt{2}}(|Bi,p_{\alpha}\rangle \mp |Bi^{'},p_{\alpha}\rangle),
 \end{eqnarray*}
 where Na(Bi) and Na$^{'}$(Bi$^{'}$) are related by inversion
 symmetry. The superscript $\pm$ indicates the parity, and $\alpha$ is
 $p_{x}$, $p_{y}$, or $p_{z}$. The bonding and anitbonding splittings
 of these states can be easily seen from the band structure along path
 $A$-$\Gamma$ shown in Fig. 2.

 By including the SOC effect in the above atomic picture, spin and orbital
 angular momentum are coupled and the new eigenstates with definite
 total angular momentum can be written as
 $|S^{\pm}_\frac{1}{2},\pm\frac{1}{2}\rangle$,
 $|P^{\pm}_{\frac{3}{2}},\pm\frac{3}{2}\rangle$,
 $|P^{\pm}_{\frac{3}{2}},\pm\frac{1}{2}\rangle$,
 $|P^{\pm}_{\frac{1}{2}}, \pm\frac{1}{2}\rangle$, where the subscript
 indicates the total angular momentum $J$. Different from the case
 with Zinc-blende structure (such as HgTe), here the heavy-hole state
 $|P^{\pm}_{\frac{3}{2}},\pm\frac{3}{2}\rangle$ and light-hole states
 $ |P^{\pm}_{\frac{3}{2}}, \pm\frac{1}{2}\rangle$ are no longer
 degenerated (with the former being higher) at the $\Gamma$ point, because Bi
 atoms are sandwiched by Na(2) atoms along the $z$ axis, and the Bi $p_z$
 orbital is lower than $p_{x,y}$ orbitals.  Furthermore, under the
 $D^4_{6h}$ symmetry, the light-hole state $|P^{\pm}_{\frac{3}{2}},
 \pm\frac{1}{2}\rangle$ and split-off state $|P^{\pm}_{\frac{1}{2}},
 \pm\frac{1}{2}\rangle$ will mix further to form the new eigen states:
 $|P_{+}^\pm,\pm\frac{1}{2}\rangle$ and
 $|P_{-}^\pm,\pm\frac{1}{2}\rangle$~\cite{cxliu}.  Nevertheless these
 mixed states are not relevant to our discussions for the 3D Dirac
 points.  The band inversion and their crossings along $\Gamma$-$A$
 line can be described by the four states
 $|S^{+}_\frac{1}{2},\pm\frac{1}{2}\rangle$ and
 $|P_{\frac{3}{2}}^{-},\pm\frac{3}{2}\rangle$.  Different from
 Bi$_2$Se$_3$,~\cite{cxliu} where all four bases have the same
 $|J_z|$=$\frac{1}{2}$, here we have two different values of
 $\frac{1}{2}$ and $\frac{3}{2}$, respectively.  This difference is
 essential to the existence and stability of 3D Dirac points observed
 here.

 Therefore, an effective 4$\times$4 $k\cdot p$ Hamiltonian using these
 four states as bases (in the order of
 $|S_\frac{1}{2}^+,\frac{1}{2}\rangle$,
 $|P^-_{\frac{3}{2}},\frac{3}{2}\rangle$,
 $|S_\frac{1}{2}^+,-\frac{1}{2}\rangle$,
 $|P^-_{\frac{3}{2}},-\frac{3}{2}\rangle$) can be constructed on
 considering the time reversal, inversion, and $D_{6h}^4$
 symmetries. The leading order Hamiltonian around $\Gamma$ reads:

\begin{eqnarray*}
  H_\Gamma(\bf{k}) & = & \epsilon_0(\bf{k})+\left(\begin{array}{cccc}
      M(\bf{k}) & Ak_{+} & 0 & B^*(\bf{k})\\
      Ak_{-} & -M(\bf{k}) & B^*(\bf{k}) & 0 \\
      0 &B(\bf{k}) & M(\bf{k}) & -Ak_{-}\\
      B(\bf{k}) & 0 & -Ak_{+} & -M(\bf{k})
\end{array}\right)
\end{eqnarray*}
where $\epsilon_0({\bf k})=C_{0}+C_{1}k_{z}^{2}+C_{2}(k_x^{2}+k_y^2)$,
$k_{\pm}=k_{x}\pm ik_{y}$, and $M({\bf
  k})=M_{0}-M_{1}k_{z}^{2}-M_{2}(k_x^{2}+k_y^2)$ with parameters $M_0,
M_1, M_2<0$ to reproduce band inversion.  By fitting the energy
spectrum of the effective Hamiltonian with that of the ab-initio
calculation, the parameters in the effective model can be
determined. For Na$_3$Bi, our fitting leads to $C_{0}$=-0.06382 eV,
$C_{1}$=8.7536 eV\AA$^2$, $C_{2}$=-8.4008 eV\AA$^2$, $M_0$=-0.08686
eV, $M_1$=-10.6424 eV\AA$^2$, $M_2$=-10.3610 eV\AA$^2$, and $A$=2.4598
eV\AA.  Please note the leading-order term of off-diagonal elements
$B({\bf k})$ has to take the high order form of $B_3k_zk_+^2$ under
the three fold-rotational symmetry and the opposite parity of
$|S\rangle$ and $|P\rangle$ states.  Evaluating the eigen values
$E({\bf k})=\epsilon_0({\bf k})\pm\sqrt{M({\bf
    k})^{2}+A^{2}k_+k_-+|B({\bf k})|^2}$, we get two gapless solutions
at {\bf k}$^c$=(0, 0, $k_z^c=\pm\sqrt{\frac{M_{0}}{M_{1}}}$), which
are the two Dirac points discussed above, which are separated along
the $\Gamma -A$ line.

If we only concentrate on the neighborhood of each crossing points
${\bf k}^c$, and neglect the high order terms (i.e., $B({\bf
  k})\approx 0$), the linearized Hamiltonian is nothing but 3D
massless Dirac fermions. The block diagonal form allows us to decouple
the 4$\times$4 matrix into two 2$\times$2 matrices, which are Weyl
fermions with degenerate energy but opposite chirality.~\cite{wan,
  HgCrSe, multilayerTRB, nodalsemimetal} The breaking of three-fold
rotational symmetry, however, will introduce a linear leading order
term of $B({\bf k})$, i. e., $B({\bf k})$=$B_1k_z$. In such a case, two
Weyl fermions will be coupled together, resulting in massive Dirac
fermions with gap, similar to the case of Bi$_2$Se$_3$ or
Bi$_2$Te$_3$.~\cite{zhangnphy, cxliu} Nevertheless, as long as the
three-fold rotational symmetry survives, the Dirac points here should
be stable and protected.

\begin{figure}[tbp]
  \includegraphics[height=144mm]{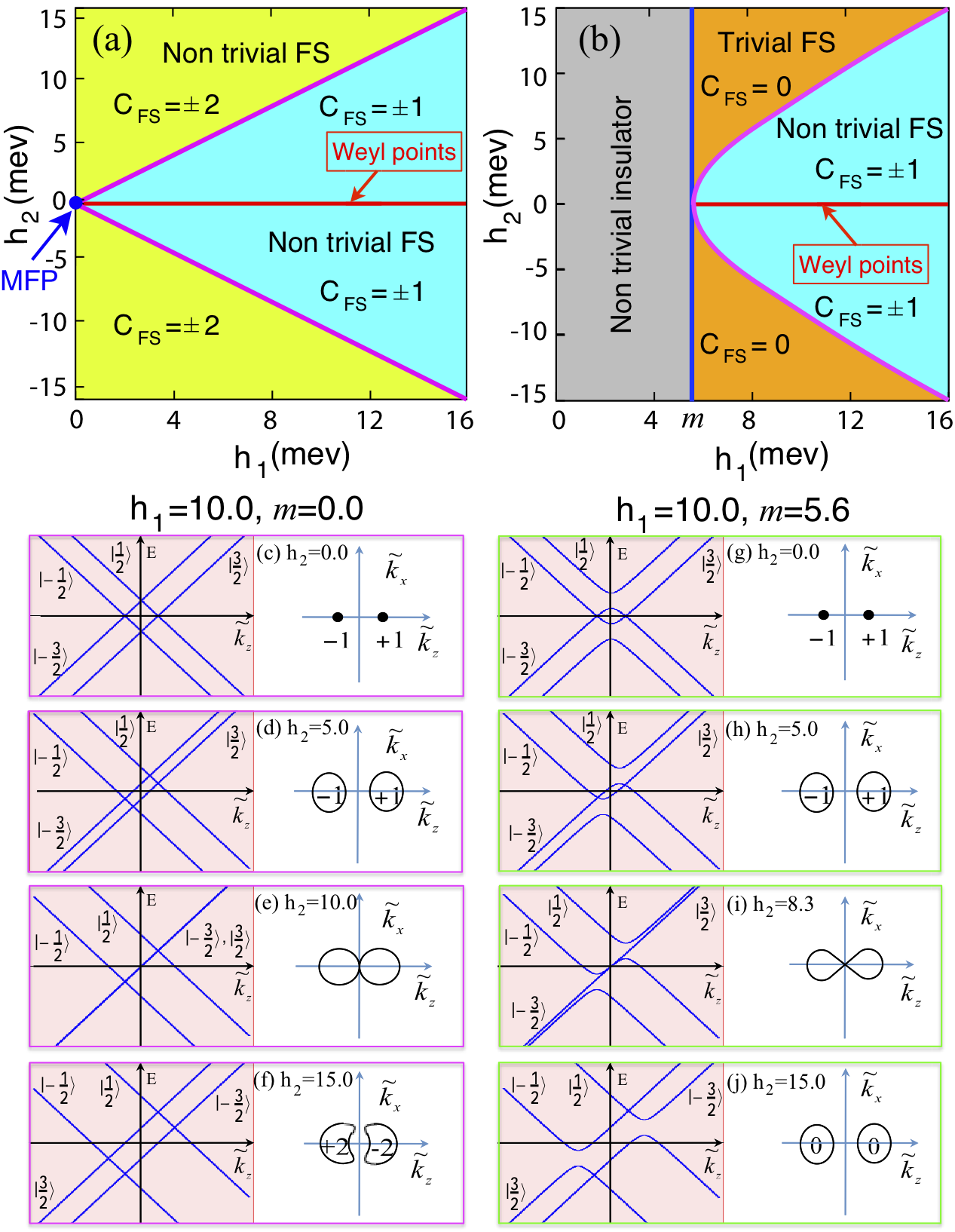}
  \caption{(Color online) Phase diagrams of Na$_3$Bi with mass term
    $m$ = 0 meV (left panels) and $m$ = 5.6 meV (right panels).  The high order
    term of $B({\bf k})$ is neglected for the case of $m$ = 5.6 meV. (a)
    and (b) Phase diagrams,  (c)-(j) Band dispersions,
    corresponding Fermi surfaces and its topological charges for some
    characteristic phases (with $h_1$ fixed to be 10.0 meV).  Only the
    neighborhood around one of the Dirac points is shown with the ${\bf
      \widetilde{k}}$ defined as ${\bf k}-{\bf k}^c$.  The
    $|\pm\frac{1}{2}\rangle$, $|\pm\frac{3}{2}\rangle$ are
    abbreviations (i.e. $J_z$ values) for the four bases, which are
    used to indicate the main component of the wave-functions for the
    states away from band crossings.  (see Sec. III.B for
    details).}
\end{figure}

\subsection{Phase diagram and topological phase transitions}

Na$_3$Bi with 3D Dirac points (i.e. the MFP state) is just located at the
phase boundary and may be driven into various topologically distinct
states by explicit breaking of
symmetries.~\cite{standardmodel,volovik-book} For simplicity of
illustration, here we focus on the effects of exchange interaction. In
general, it can be induced by magnetic doping as in diluted magnetic
semiconductors~\cite{exchange} or by an external field. Other
symmetry-breaking terms (such as inversion, mirror or two fold
rotational symmetries of the crystal) may play the similar roles and
can be analyzed analogously.  Since the $|S\rangle$ and $|P\rangle$
are different orbitals (or pseudo spins), we may in general separate
any exchange splitting into orbital-dependent and orbital-independent parts as
$H_{ex1}=h_1 \sigma_z\otimes\tau_z$ and $H_{ex2}=h_2 \sigma_z\otimes
I$, where $h_1$ and $h_2$ are field strengths (along the $z$ direction),
and $\vec{\sigma}$ and $\vec{\tau}$ are Pauli matrices describing spin
and pseudo-spin respectively. The total Hamiltonian is given as
$H=H_\Gamma+H_{ex1}+H_{ex2}$, and the resulting phase diagram is shown
in Fig.4.

If the three fold rotational symmetry of crystal is preserved
(i.e. $B({\bf k})$=$B_3k_z^ck_+^2$, see left panels of Fig.4),
starting from the MFP state ($h_1$=0, $h_2$=0), the state with Weyl
points will be introduced by $h_1$ (the horizontal axis), because such an
exchange field will split the Dirac point into two separated Weyl
points in momentum space. If the $h_2$ is further introduced, however,
the two Weyl points will separate energetically, and a system with
fermi surfaces (FS) will be obtained. On the other hand, if the
three-fold rotational symmetry is broken, a mass term $m\approx
B_1k_z^c$ will be induced as the leading-order term of $B({\bf k})$.
For example, $m$ can be estimated to be the gap size (5.6 meV) at {\bf
  k}$^c$ when 1\% compression along the $y$ axis is applied. Then the high
order term of $B({\bf k})$ can be neglected, and a topologically
non-trivial insulating phase is obtained (see right panels of Fig. 4). In
such a case, the Weyl semimetal phase can be driven only when $h_1$ is
larger than the mass term $m$.

The FS states can be further classified according to the topological
charge (or Chern number $C_{FS}$) enclosed by the FS. $C_{FS}$ is
defined as the net flux of the Berry phase gauge field penetrating the
fermi surface,
\begin{eqnarray*}
  C_{FS}=\frac{1}{2\pi}\int_{FS}  ({\bf \nabla}_{\bf
    k}\times {\bf A(k)})    \cdot d{\bf S}
\end{eqnarray*}
where the integrand is the Berry curvature, {\bf A(k)}=$-i\langle
u_{\bf k}|\nabla_{\bf k}|u_{\bf k}\rangle$ is the adiabatic Berry
connection for the states $|u_{\bf k}\rangle$ at the fermi level, and
$d{\bf S}$ points from low to high energy. For the case $m$=5.6meV
(right panels), the two distinct FS states, trivial ($C_{FS}=0$) and
non-trivial ($C_{FS}=\pm1$), are separated by the line defined as
$h_1^2-h_2^2$=$m^2$. If $m=0$ (left panels), both FS states are
non-trivial, but with different topological charges ($C_{FS}=\pm 1$ or
$\pm 2$). The appearance of the $C_{FS}$=$\pm 2$ phase in this case is due
to the $B_3k_zk_+^2$ term of $B({\bf k})$.  At the boundary between
distinct FS states, the fermi surface spheres should be connected, and
the $C_{FS}$ becomes ill-defined.  The non-trivial FS states may
become important for the topological superconductivity.~\cite{xlqi-sc}

\subsection{Expected distinct physical properties}

Even without the exchange splitting, we can expect some particular
physical properties for such compounds. First of all, we will
expect the quantum spin Hall effect in $z$-oriented Na$_3$Bi thin film (or a
Na$_3$Bi/Na$_3$Sb quantum well). Due to the quantum size effect, the
$k_z$ is further quantized, and in general the 2D band structures of
Na$_3$Bi thin film will be fully gapped. Then depending on the number
of band inversions associated with the subbands, the system should
cross over between trivial and non-trivial 2D insulators oscillatorily
as a function of film thickness.~\cite{cxliu-osci}  Our estimated
first critical thickness of Na$_3$Bi is 35~\AA, below (above) which
the film is a trivial (non-trivial) insulator.  Second, we will
expect giant diamagnetism of the 3D massless Dirac fermion.~\cite{diamag}
The diamagnetic susceptibility,
$\chi(\varepsilon)$$\sim$log$\frac{1}{\varepsilon}$, should diverge
logarithmically when the chemical potential approaches the 3D Dirac
points (i.e. $\varepsilon$$\sim$0),~\cite{diamag} much stronger than
that in narrow gap semimetals like Bismuth. In fact, early
experimental measurement had found distinct diamagnetism in the $A_3$Bi
system.\cite{na3bidiamag} Third, we will also expect linear quantum
magnetoresistance (MR) as proposed by
A. A. Abrikosov.~\cite{quantum-MR} In conventional metals with closed
fermi surface, the MR should behavior quadratically at a low field and
saturate at a high field. However, for the 3D massless Dirac fermionic
gas, the MR will have linear field dependence if only the lowest
Landau level is occupied. This idea has been examined for the
Ag$_2$Te both experimentally and
theoretically,~\cite{quantum-MR,Ag2Te} where the Dirac-type energy
dispersion is not obvious. Having Na$_3$Bi with 3D Dirac points, it
will be straightforward to check the quantum MR
proposal.~\cite{quantum-MR}

\section{Conclusion} \label{Conclusion}

In summary, based on the first-principles calculations and effective
model analysis, we have shown that the long-pursuing examples with bulk
3D Dirac points can be actually realized in existing compounds $A_3$Bi
($A$=Na, K, Rb). It is important to note that this state and its Dirac
points are protected by crystal symmetry, and therefore stable. We
have demonstrated that this state is located at the topological phase
boundary, and can be driven into various topologically distinct
phases, such as topological insulator, topological metal (with
non-trivial Fermi surfaces), and Weyl semimetal states, by explicit
breaking of symmetries. It therefore may provide us a
condensed-matter simulator of the Standard model, from the view point of an
emerging relativistic quantum field at low energy. In addition, we
have shown that the state itself is unconventional in the sense that
it shows Fermi arcs on the surface, giant diamagnetism and linear
quantum magnetoresistance in the bulk, and quantum spin Hall effect in
the quantum-well or thin-film structure. Experimenters are strongly
encouaged to test those proposals and phenomena.

\begin{acknowledgments}
We acknowledge the supports from NSF of China, the 973 program of
China, and the Hundred Talents Project of the Chinese Academy of
Sciences.
\end{acknowledgments}


\begin{references}


\bibitem{standardmodel} F. R. Klinkhamer, G. E. Volovik,
  Int. J. Mod. Phys. A~{\bf 20}, 2795 (2005).


\bibitem{volovik-book} Grigory E. Volovik, {\it The Universe in a
    Helium Droplet} (Clarendon Press, Oxford), 2003.


\bibitem{TIreview} M. Z. Hasan, and C. L. Kane, Rev. Mod. Phys. {\bf
    82}, 3045 (2010).

\bibitem{TIreview-2} X.-L. Qi, and S.-C. Zhang, Rev. Mod. Phys. {\bf
    83}, 1057 (2011).


\bibitem{TIclass-1} A. P. Schnyder, S. Ryu, A. Furusaki,
  A. W. W. Ludwig, Phys. Rev. B {\bf 78}, 195125 (2008).




\bibitem{zhangnphy} H. J. Zhang, C. X. Liu, X. L. Qi, X. Dai, Z. Fang,
  S. C. Zhang, Nature Phys. {\bf 5}, 438 (2009).

\bibitem{Xia} Y. Xia, D. Qian, D. Hsieh, L. Wray, A. Pal, H. Lin,
  A. Bansil, D. Grauer, Y. S. Hor, R. J. Cava, and M. Z. Hasan, Nature
  Phys. {\bf 5}, 398 (2009).


\bibitem{Chen} Y. L. Chen, J. G. Analytis, J.-H. Chu, Z. K. Liu, S.-K.
  Mo, X. L. Qi, H. J. Zhang, D. H. Lu, X. Dai, Z. Fang, S. C. Zhang,
  I. R. Fisher, Z. Hussain, Z.-X. Shen, Science {\bf 325}, 178 (2009).



\bibitem{He3} Anthony J. Leggett, Rev. Mod. Phys. {\bf 47}, 331
  (1975).


\bibitem{wan}
  X. G. Wan, A. M. Turner, A. Vishwanath, S. Y. Savrasov, Phys. Rev. B
  {\bf 83}, 205101 (2011).

\bibitem{HgCrSe}
  G. Xu, H. M. Weng, Z. J. Wang, X. Dai, Z. Fang,
  Phys. Rev. Lett. {\bf 107}, 186806 (2011).

\bibitem{multilayerTRB}
A. A. Burkov, L. Balents, Phys. Rev. Lett. {\bf 107}, 127205 (2011).


\bibitem{nodalsemimetal}
  A. A. Burkov, M. D. Hook, L. Balents, Phys. Rev. B {\bf 84}, 235126
  (2011).


\bibitem{Murakami} S. Murakami, New J. Phys. {\bf 9}, 356 (2007).

\bibitem{ZhangW} W. Zhang, R. Yu, H. J. Zhang, X. Dai, Z. Fang, New
  J. Phys. {\bf 12}, 065013 (2010).


\bibitem{Hasan-phaseT} S. Y. Xu, Y. Xia, L. A. Wray, S. Jia, F. Meier,
  J. H. Dil, J. Osterwalder, B. Slomski, A. Bansil, H. Lin,
  R. J. Cava, M. Z. Hasan, Science {\bf 332}, 560 (2011).


\bibitem{Ando-phaseT} T. Sato, K. Segawa, K. Kosaka, S. Souma,
  K. Nakayama, K. Eto, T. Minami, Y. Ando, Nature Phys. {\bf 7}, 840
  (2011).








\bibitem{BiO2} S. M. Young, S. Zaheer, J. C. Y. Teo, C. L. Kane,
  E. J. Mele, A. M. Rappe, Phys. Rev. Lett. {\bf108}, 140405 (2012).





\bibitem{A3Sb-zunger} S. H. Wei, A. Zunger, Phys. Rev. B {\bf 35},
  3952 (1987).


\bibitem{A3Bi-elec} M. Tegze, J. Hafner, J. Phys.: Cond. Matt. {\bf
    4}, 2449 (1992).


\bibitem{A3B-struc} T. B. Massalski, {\it Binary Alloy Phase
    Diagrams}, (ASM International, Materials Park, Ohio), 1990.


\bibitem{Na3Bi-struc} G. Brauer, E. Zintl, Z. Phys. Chem., Abt. B {\bf
    37}, 323 (1937).





\bibitem{Na3Sb}
A. R. H. F. Ettema, and R. A. de Groot, Phys. Rev. B {\bf 61}, 10035 (2000).

\bibitem{HSE06} J. Heyd, G. E. Scuseria, and M. Ernzerhof, J. Chem. Phys. \textbf{124}, 219906 (2006).


\bibitem{K3Sb} A. R. H. F. Ettema, R. A. de Groot, J. Phys.:
  Cond. Matt. {\bf 11}, 759 (1999).









\bibitem{Kino} H. Kino, T. Miyazaki, J. Phys. Soc. Jpn. {\bf 75},
  034704 (2006).



\bibitem{Pickett} V. Pardo, W. E. Pickett, Phys. Rev. Lett. {\bf 102},
  166803 (2009).

\bibitem{cxliu} C. X. Liu, X. L. Qi, H. J. Zhang, X. Dai, Z. Fang,
  S. C. Zhang, Phys. Rev. B {\bf 82}, 045122 (2010).

\bibitem{exchange} R. Yu, W. Zhang, H. J. Zhang, S. C. Zhang,
 X. Dai and Z. Fang, Science {\bf 329}, 61 (2010).


\bibitem{xlqi-sc} X. L. Qi, T. L. Hughes, S. C. Zhang, Phys. Rev. B
  {\bf 81}, 134508 (2010).


\bibitem{cxliu-osci} C. X. Liu, H. J. Zhang, B. H. Yan, X. L. Qi,
  T. Frauenheim, X. Dai, Z. Fang, S. C. Zhang, Phys. Rev. B {\bf 81},
  041307 (2010).



\bibitem{diamag} M. Koshino, T. Ando, Phys. Rev. B {\bf 81}, 195431
  (2010); and see references therein.

\bibitem{na3bidiamag} E. R\"ober, K. Hackstein, H. Coufal, S. Sotier, Phys. Status Solidi (B), {\bf 93}, K99 (1979);
K. Hackstein, S. Sotier and E. L\"uscher,
J. Phys. Colloques 41 C8-49 (1980).

\bibitem{quantum-MR} A. A. Abrikosov, Phys. Rev. B {\bf 58}, 2788
  (1998); and see references therein.

\bibitem{Ag2Te} W. Zhang, R. Yu, W. Feng, Y. Yao, H. M. Weng, X. Dai,
  Z. Fang, Phys. Rev. Lett. {\bf 106}, 156808 (2011).






\end{references}
\end{document}